# Bounds and Inequalities Relating $h$-Index, $g$-Index, $e$-Index and Generalized Impact Factor

Ash Mohammad Abbas
Department of Computer Engineering
Aligarh Muslim University
Aligarh - 202002, India
am.abbas.ce@amu.ac.in

October 28, 2018

**Abstract**

Finding relationships among different indices such as $h$-index, $g$-index, $e$-index, and generalized impact factor is a challenging task. In this paper, we describe some bounds and inequalities relating $h$-index, $g$-index, $e$-index, and generalized impact factor. We derive the bounds and inequalities relating these indexing parameters from their basic definitions and without assuming any continuous model to be followed by any of them.

**Keywords:** Impact factor, $h$-index, $g$-index, $e$-index.

## 1 Introduction

A lot of research is carried out by people working in different areas. Sometimes, one needs to evaluate the quality of the research produced by individual authors or groups of authors. The quality of research produced by authors is, generally, evaluated in terms a ranking parameter which is, generally, based on the number of citations received by the papers produced by the authors. There are many types of ranking parameters presented in the literature for evaluating the quality of research such as $h$-index [12], $g$-index [5], $e$-index [14], and impact factor. Some of these parameters can also be used to evaluate the quality of research published by a journal such as $h$-index, $g$-index, $e$-index, and impact factor. The impact factor in the long term becomes the average number of citations per published paper. This long term impact factor is termed as the *generalized impact factor* in [1].

While one has computed an index for evaluating the quality of research, one would like get an indication about the other types of indices. To have such an indication, one needs to know how an index is related to other indices. The relationships among $h$-index, $g$-index, and $e$-index are described in [15]. However, in [15], the indices are assumed to follow a continuous distribution.

A relation between $h$-index and impact factor is described in [8] using a power law model called the Lotka's model. An attempt to relate $h$-index, $g$-index, and the *generalized impact factor* is made in [1]. The relationships, therein, contain only inequalities among the parts of the number of citations. The parts for the number of citations are not bounded by an expression, therefore, the relationships described therein seem to be trivial. However, it was an attempt to derive the relationships among the indices using their basic definitions and no distribution or a continuous model was assumed for any of the index.

In this paper, we describe the bounds for the $h$-index and $g$-index in terms of the indices and the *generalized impact factor*. We derive these bounds from the very basic definitions of the indices and the *generalized impact factor* without assuming any model or any continuous distribution to be followed by any of these indices.

In what follows, we provide a brief overview of the indices and the *generalized impact factor*.

## 2 Overview of Indices and Impact Factor

In this section, we present an overview of impact factor and different types of indices.

### 2.1 The $h$-Index

Suppose the papers are arranged in descending order of the number of citations. Let $c_i$ be the number of citations of a paper numbered $i$. The $h$-index [12], when papers are arranged in descending number of their citations, can be defined as follows.

$$h = \max(i) : c_i \geq i. \qquad (1)$$

By definition, $h$-index is the largest number, $h$, such that the papers arranged in their decreasing order of citations have at least $h$ number of citations.

### 2.2 The $g$-Index

According to the definition of $g$-index, if the papers are arranged in the descending order of their number of citations, $g$ is the largest number such that the summation of the number of citations is at least $g^2$. In other words, when papers are arranged in descending order of their citations, $g$-index can be defined as follows.

$$g = \max(i) : \sum_i c_i \geq i^2. \qquad (2)$$

Note that $g$-index is the largest number $i$ such that $\sum_i c_i \geq i^2$.

## 2.3 The $e$-Index

The $e$-index is defined in [14] to serve as a complement for the $h$-index. The definition of $e$-index is as follows.

$$\begin{aligned} e^2 &= \sum_{i=1}^{h}(c_i - h) \\ &= \left(\sum_{i=1}^{h} c_i\right) - h^2. \end{aligned} \tag{3}$$

Alternatively, (3) can be written as follows.

$$\sum_{i=1}^{h} c_i = h^2 + e^2. \tag{4}$$

*Remark:* In the definitions of $h$-index (as given by (1)) and that of $g$-index (as given by (2)), we have intentionally ignored the time $T$ at which we are considering their values. This is done to keep their definitions simple, and defining so there is no loss of generality as far as the discussion in this work is concerned. For precise definitions of the indices incorporating the time, one is referred to [3]. The same is true for the $e$-index.

## 2.4 Generalized Impact Factor

Let $c_i$ be the number of citations of $i$th paper published in a journal. Let there be total $P$ papers published by the journal. The *generalized impact factor* [1][1], which should better be called as the *average number of citations per paper* [1] or an *impact factor without a time window constraint* [1] or an *impact factor with a time frame from the beginning of the publication of the journal till now or till the time of evaluation* [1] or simply an *impact factor* [8] is defined as follows.

$$\begin{aligned} I_f &= \frac{\sum_{i=1}^{P} c_i}{P} \\ &= \frac{C}{P}. \end{aligned} \tag{5}$$

# 3 Analysis of Relationships

In this section, we describe how indices and generalized impact factor are related to one another.

---

[1]A better name for the *generalized impact factor* is *average number of citations per paper* or *long term impact factor*.

## 3.1 Impact Factor, $h$-Index and $e$-Index

We state the following theorem that relates these parameters.

**Theorem 1.** *Let $P$ be the number of papers published in a journal. The $h$-index, $e$-Index and impact factor are related by the following inequality.*

$$h \geq \left\lfloor I_f - \frac{e^2}{P} \right\rfloor. \tag{6}$$

*Proof.* Using (5), the total number of citations of a journal can be written as follows.

$$\sum_{i=1}^{P} c_i = I_f P. \tag{7}$$

The citations appearing in the L.H.S. of (7) can be broken into two parts, one from 1 to $h$ and the other from $h+1$ to $P$, as given below.

$$\sum_{i=1}^{h} c_i + \sum_{i=h+1}^{P} c_i = I_f P. \tag{8}$$

Using (8) and (4), we have,

$$h^2 + e^2 + \sum_{i=h+1}^{P} c_i = I_f P. \tag{9}$$

Now, we have,

$$\begin{aligned} c_{h+1} &\leq h \\ c_{h+2} &\leq h \\ \ldots &\leq \ldots \\ c_P &\leq h. \end{aligned} \tag{10}$$

Therefore, we have,

$$\sum_{i=h+1}^{P} c_i \leq (P-h)h. \tag{11}$$

Using (11) and (9), we have,

$$\begin{aligned} h^2 + e^2 + (P-h)h &\leq I_f P \\ e^2 + Ph &\leq I_f P. \end{aligned} \tag{12}$$

In other words, we have,

$$h \geq I_f - \frac{e^2}{P}. \tag{13}$$

Since $h$ is a whole number, therefore, we can write,

$$h \geq \left\lfloor I_f - \frac{e^2}{P} \right\rfloor.$$

□

In other words, we can say that

$$h = \Omega\left(\left\lfloor I_f - \frac{e^2}{P}\right\rfloor\right) \tag{14}$$

where, $\Omega$ denotes the lower bound.

### 3.2 The $g$-Index, $h$-Index, and $e$-Index

We state the following theorem that provides an inequality relating these indices.

**Theorem 2.** *The h-index, g-index, and e-index are related with the following inequality.*

$$h \geq \left\lfloor g - \frac{e^2}{g}\right\rfloor. \tag{15}$$

*Proof.* Let the the papers are arranged in the descending order of their citations. From the definition of $g$-index, as given in (2), we have,

$$g = \max(i) : \sum_i c_i \geq i^2. \tag{16}$$

At $i = g$, we have,

$$\sum_{i=1}^{g} c_i \geq g^2. \tag{17}$$

Breaking the number of citations in the L.H.S. of (17) into parts, we have,

$$\sum_{i=1}^{h} c_i + \sum_{i=h+1}^{g} c_i \geq g^2. \tag{18}$$

From (4) and (18), we have,

$$h^2 + e^2 + \sum_{i=h+1}^{g} c_i \geq g^2. \tag{19}$$

In other words,

$$g^2 - (h^2 + e^2) \leq \sum_{i=h+1}^{g} c_i. \tag{20}$$

Now, we have,

$$\begin{aligned} c_{h+1} &\leq h \\ c_{h+2} &\leq h \\ \dots &\leq \dots \\ c_g &\leq h. \end{aligned} \tag{21}$$

Therefore, we have,
$$\sum_{i=h+1}^{g} c_i \leq (g-h)h. \tag{22}$$

Using (20) and (22), we have,
$$g^2 - (h^2 + e^2) \leq (g-h)h. \tag{23}$$

Or,
$$g^2 - e^2 \leq gh. \tag{24}$$

Rearranging (24), we have,
$$h \geq g - \frac{e^2}{g}. \tag{25}$$

Since all these indices, $h$, $g$, and $e$ are integers, therefore, (25) can be written as follows.
$$h \geq \left\lfloor g - \frac{e^2}{g} \right\rfloor.$$

□

In other words, Theorem 2 provides a lower bound for $h$-index in terms of the $g$-index and the $e$-index.
$$h = \Omega\left(\left\lfloor g - \frac{e^2}{g} \right\rfloor\right). \tag{26}$$

We have the following lemma that provides a bound for the $g$-index.

**Lemma 1.** *An upper bound for $g$-index is as follows.*
$$g = O\left(\left\lceil h + \frac{e^2}{h} \right\rceil\right). \tag{27}$$

*Proof.* From (20), we have,
$$g^2 - (h^2 + e^2) \leq \sum_{i=h+1}^{g} c_i.$$

In (21), if we put $g$ at the R.H.S. for $h+1 \leq i \leq g$, $c_i \leq g$, we get,
$$\sum_{i=h+1}^{g} c_i \leq (g-h)g. \tag{28}$$

Therefore, from (20), we have,
$$g^2 - (h^2 + e^2) \leq (g-h)g. \tag{29}$$

Or,
$$h^2 + e^2 - gh \geq 0. \tag{30}$$
Or,
$$h^2 + e^2 \geq gh. \tag{31}$$
This gives us,
$$g \leq h + \frac{e^2}{h}. \tag{32}$$
Again, all these indices are whole numbers, therefore, we can write,
$$g \leq \left\lceil h + \frac{e^2}{h} \right\rceil. \tag{33}$$
Alternatively,
$$g = O\left(\left\lceil h + \frac{e^2}{h} \right\rceil\right).$$
□

We now prove another theorem that provides an upper bound for the $g$-index in terms of $h$-index and $e$-index.

**Theorem 3.** *An upper bound for g-index in terms of h-index and e-index is as follows.*
$$g = O(h + e). \tag{34}$$

*Proof.* Using (22), we have,
$$g^2 - gh - e^2 \leq 0. \tag{35}$$
This resembles to the quadratic equation $ax^2 + bx + c = 0$, whose roots are as follows.
$$r_i \Big|_{i=1}^{2} = \frac{-b \pm \sqrt{b^2 - 4ac}}{2a}.$$
Here, we have, $a = 1$, $b = -h$, $c = -e^2$, therefore, the only root for $g$-index is,
$$g \leq \frac{h + \sqrt{h^2 + 4e^2}}{2}. \tag{36}$$
Now, we know that $(h + 2e)^2 = h^2 + 4e^2 + 4eh$. In other words, we have,
$$h + 4e^2 \leq (h + 2e)^2. \tag{37}$$
This implies that
$$\sqrt{h^2 + 4e^2} \leq h + 2e. \tag{38}$$
Using (36) and (38), we have,
$$\begin{aligned} g &\leq \frac{h + (h + 2e)}{2} \\ &\leq h + e. \end{aligned} \tag{39}$$
In other words, $g = O(h + e)$. □

## 3.3 The $h$-Index, $g$-Index, and Impact Factor

We state the following theorem that relates these parameters.

**Theorem 4.** *The generalized impact factor, g-index, and h-index are related as per the following inequality.*

$$h \geq \left\lfloor \frac{I_f P - g^2}{P - g} \right\rfloor. \tag{40}$$

*Proof.* From (5), we have,

$$\sum_{i=1}^{P} c_i = I_f P. \tag{41}$$

Breaking the number of citations in the L.H.S. of (41), we have,

$$\sum_{i=1}^{g} c_i + \sum_{i=g+1}^{P} c_i = I_f P. \tag{42}$$

From (17), we have, $\sum_{i=1}^{g} c_i \geq g^2$. However, we need to point out that at the point $i = g, \sum_{i=1}^{g} c_i$ just exceeds $g^2$. The amount by which $\sum_{i=1}^{g} c_i$ may exceed $g^2$ depends upon the value of $c_i$ at $i = g$, i.e. $c_g$. Therefore, for all practical purposes, we can assume $\sum_{i=1}^{g} c_i \approx g^2$. As a result, we have,

$$g^2 + \sum_{i=g+1}^{P} c_i \approx I_f P. \tag{43}$$

Now, we have,

$$\begin{aligned} c_{g+1} &\leq h \\ c_{g+2} &\leq h \\ \ldots &\leq \ldots \\ c_P &\leq h. \end{aligned} \tag{44}$$

Therefore, we have,

$$\sum_{i=g+1}^{P} c_i \leq (P - g)h. \tag{45}$$

Using (43) and (44), we have,

$$g^2 + (P - g)h \geq I_f P. \tag{46}$$

Or,

$$(P - g)h \geq I_f P - g^2. \tag{47}$$

As a result, we get,

$$h \geq \frac{I_f P - g^2}{P - g}. \tag{48}$$

Since $h$ is a whole number, therefore, (48)

$$h \geq \left\lfloor \frac{I_f P - g^2}{P - g} \right\rfloor.$$

□

In other words, Theorem 4 states another lower bound for the $h$-index which is as follows.

$$h = \Omega\left(\left\lfloor \frac{I_f P - g^2}{P - g} \right\rfloor\right). \tag{49}$$

While proving Theorem 4, we come across the following observation. We could have taken on the R.H.S. of (44), the value of $g$-index rather than $h$-index. We have taken the value of $h$-index to make the lower bound on $h$-index to be tight enough. To understand it better, let us take the value of $g$-index in (44). We now state a lemma that gives a lower bound (though loose) on the $g$-index.

**Lemma 2.** *A loose lower bound on the $g$-index is as follows.*

$$g = \omega(I_f). \tag{50}$$

*Proof.* In other words, we could have written (44) as follows.

$$\begin{aligned} c_{g+1} &\leq g \\ c_{g+2} &\leq g \\ \ldots &\leq \ldots \\ c_P &\leq g. \end{aligned} \tag{51}$$

Therefore, we have,

$$\sum_{i=g+1}^{P} c_i \leq (P-g)g. \tag{52}$$

Putting it in (43), we get,

$$g^2 + (P-g)g \geq I_f P. \tag{53}$$

Observe that the $h$-index has now vanished and does not appear anywhere in (53). From (53), we have,

$$g \geq I_f. \tag{54}$$

In other words, the impact factor can serve as a lower bound for the $g$-index.

$$g = \omega(I_f). \tag{55}$$

□

Table 1: Number of citations, indices, and generalized impact factor of authors.

| Author | $h$ | $g$ | $e^2$ | $e$ | $P$ | $I_f$ | $\sum_{i=1}^{P} c_i$ |
|---|---|---|---|---|---|---|---|
| A | 4 | 7 | 23 | 5 | 17 | 4.12 | 70 |
| B | 8 | 11 | 35 | 6 | 30 | 5.20 | 156 |
| C | 11 | 18 | 167 | 13 | 30 | 12.07 | 362 |
| D | 12 | 20 | 198 | 15 | 52 | 9.31 | 484 |
| E | 14 | 30 | 623 | 25 | 59 | 17.10 | 1009 |
| F | 28 | 43 | 750 | 28 | 181 | 14.07 | 2546 |

Table 2: Part of citations from paper numbered $(g+1)$ to $P$ and their bounds.

| Author | $\sum_{i=g+1}^{P} c_i$ | $(P-g)h$ | $g^2 + \sum_{i=g+1}^{P} c_i$ |
|---|---|---|---|
| A | 20 | 40 | 69 |
| B | 38 | 152 | 159 |
| C | 23 | 132 | 345 |
| D | 74 | 384 | 474 |
| E | 62 | 406 | 962 |
| F | 669 | 3864 | 2518 |

However, this lower bound may not be tight enough, therefore, we call it a loose lower bound, and in Lemma 2, it is denoted by the symbol small omega, $\omega$, which is generally used to denote a loose lower bound. The reason for the looseness of the lower bound is that we have assumed in the (51), $c_i \leq g$ for $(g+1) \leq i \leq P$, which is a loose assumption as compared to the assumption made in (44).

In what follows, we discuss some results to validate the analysis.

Table 3: Part of citations from paper numbered $h+1$ to $g$, and from $h+1$ to $P$ and their bounds.

| Author | $\sum_{i=h+1}^{P} c_i$ | $(P-h)h$ | $\sum_{i=h+1}^{g} c_i$ | $(g-h)h$ | $(g-h)g$ |
|---|---|---|---|---|---|
| A | 31 | 52 | 11 | 12 | 21 |
| B | 57 | 176 | 19 | 24 | 33 |
| C | 64 | 209 | 41 | 77 | 126 |
| D | 138 | 480 | 64 | 96 | 160 |
| E | 210 | 630 | 148 | 224 | 480 |
| F | 1001 | 4284 | 332 | 420 | 645 |

Table 4: Bounds on the $h$-index.

| Author | $h$-index | $I_f - \frac{e^2}{P}$ | $g - \frac{e^2}{g}$ | $\frac{I_f P - g^2}{P - g}$ |
|---|---|---|---|---|
| A | 4 | 3 | 3 | 2 |
| B | 8 | 4 | 7 | 1 |
| C | 11 | 5 | 8 | 3 |
| D | 12 | 6 | 10 | 2 |
| E | 14 | 5 | 9 | 3 |
| F | 28 | 9 | 25 | 5 |

Table 5: Bounds on the $g$-index.

| Author | $g$-index | $h + \frac{e^2}{h}$ | $h + e$ |
|---|---|---|---|
| A | 7 | 10 | 9 |
| B | 11 | 13 | 14 |
| C | 18 | 27 | 24 |
| D | 20 | 29 | 27 |
| E | 30 | 59 | 39 |
| F | 43 | 55 | 56 |

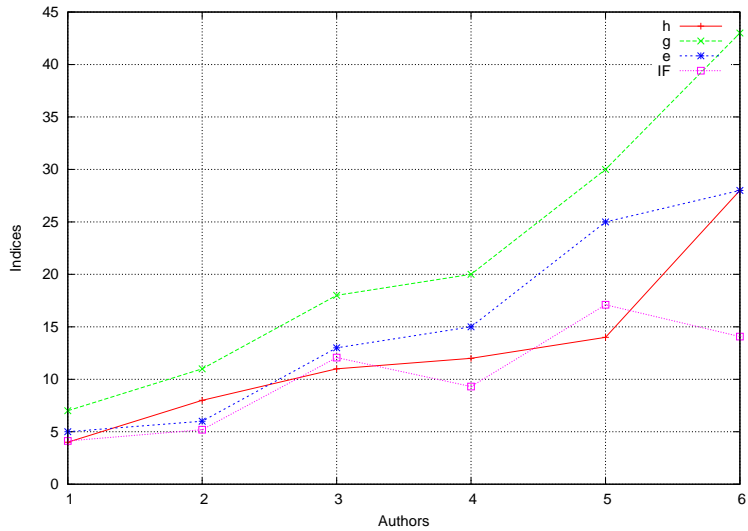

Figure 1: The $h$-index, $g$-index, $e$-index, and generalized impact factor of authors $A$ through $F$.

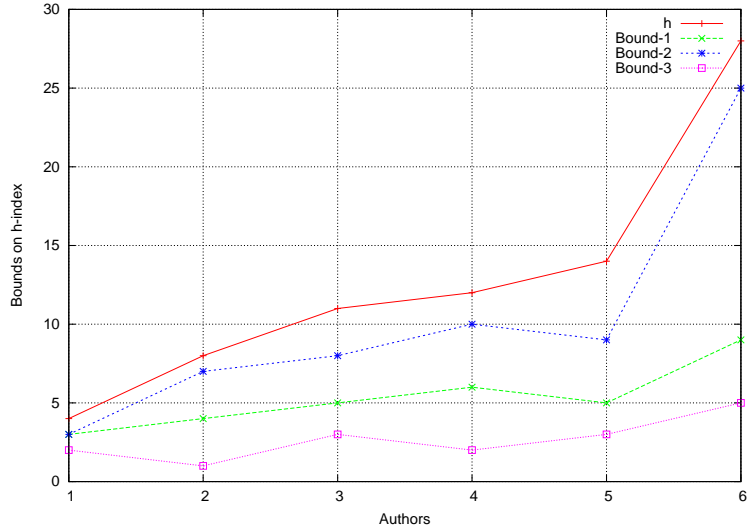

Figure 2: The $h$-index and its lower bounds given by Theorem 1, Theorem 2 and Theorem 4 for authors $A$ through $F$.

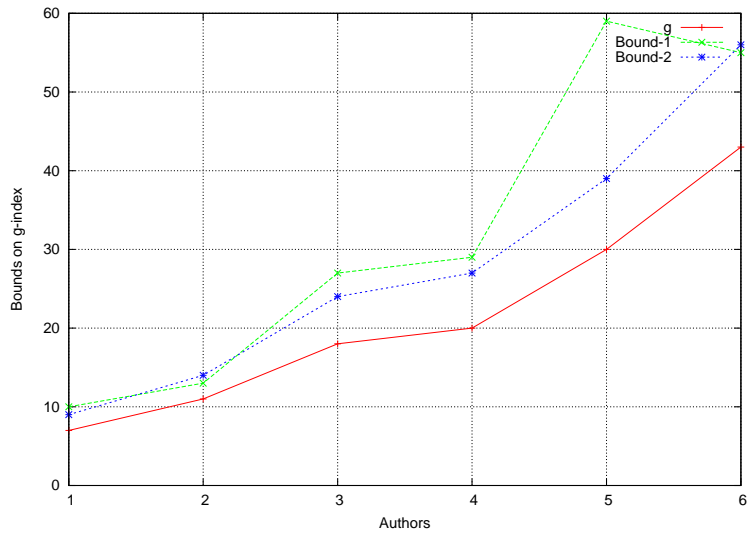

Figure 3: The $g$-index and its upper bounds given by Lemma 1 and Theorem 3 for authors $A$ through $F$.

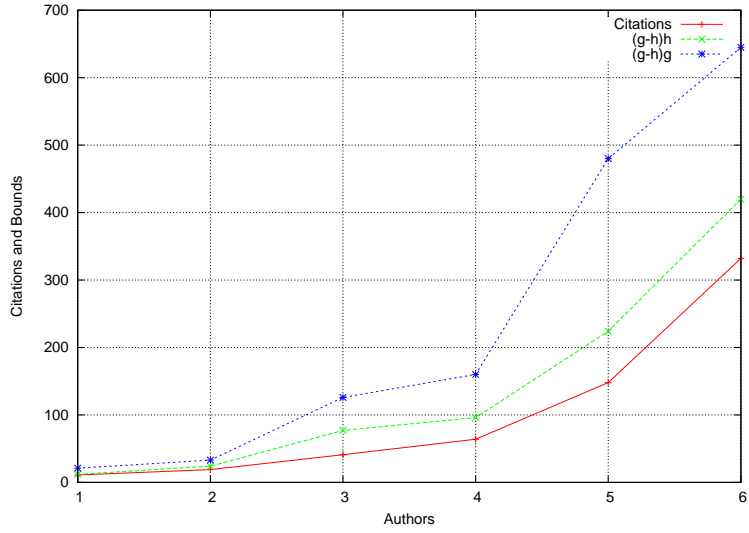

Figure 4: The number of citations, $\sum_{i=h+1}^{g} c_i$, and its upper bounds, $(g-h)h$, and $(g-h)g$, for authors $A$ through $F$.

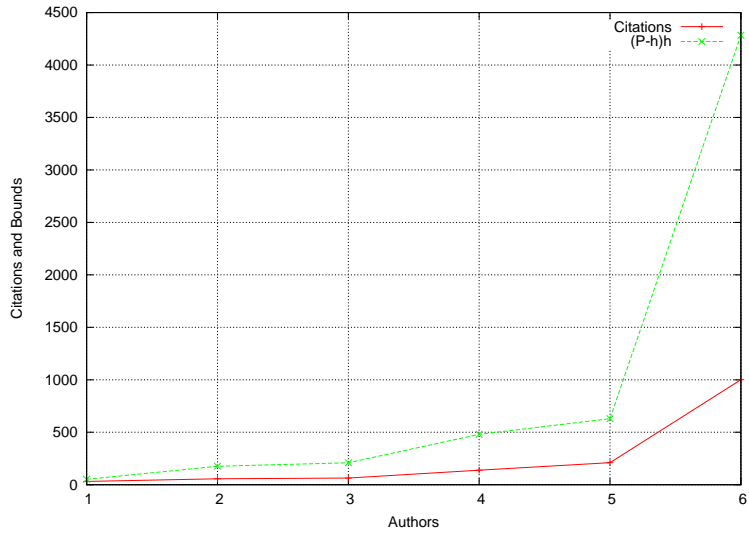

Figure 5: The number of citations, $\sum_{i=h+1}^{P} c_i$, and its upper bound, $(P-h)h$, for authors $A$ through $F$.

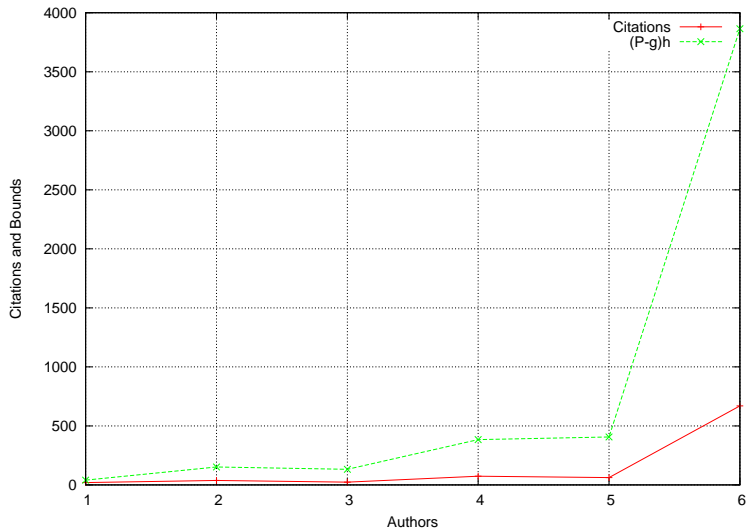

Figure 6: The number of citations, $\sum_{i=g+1}^{P} c_i$, and its upper bound, $(P - g)h$, for authors $A$ through $F$.

## 4 Results and Discussion

Table 1 provides different parameters such as the total number of citations, $h$-index, $g$-index, $e^2$, generalized impact factor for individual authors. We have selected authors from different fields[2]. Note that Table 2 and Table 3 contain the values of the number of citations in parts and the values of corresponding approximations used for these parts. These values may be useful in understanding the differences among the actual values of the indices and their corresponding bounds.

Table 4 shows bounds on the $h$-index as described in the previous section. The third column of the table represents a lower bound on $h$-index. It is clear that $h \geq \left\lfloor I_f - \frac{e^2}{P} \right\rfloor$. However, the lower bound $h = \Omega \left( \left\lfloor I_f - \frac{e^2}{P} \right\rfloor \right)$ is not very tight. The reason is that while deriving this bound, we have used an approximation $\sum_{i=h+1}^{P} c_i \leq (P - h)h$, which is not a very tight approximation. The looseness of the lower bound comes from the fact that we have assumed $c_i \leq h$ for $h + 1 \leq i \leq P$. Practically, $c_i << h$ for $h + 1 \leq i \leq P$, and that introduces the looseness.

Coming to the fourth column of Table 4 that represents a lower bound on the $h$-index given by $h \geq \left\lfloor g - \frac{e^2}{g} \right\rfloor$. Note that this lower bound is more tight as compared to the lower bound in the third column of this table. For this lower bound, whatever looseness remains is due to the fact that while deriving the lower bound we have taken an approximation $\sum_{i=h+1}^{g} c_i \leq (g-h)h$ by assuming that each $c_i \leq h$ for $(h+1) \leq i \leq g$. Practically, it may happen that $c_i << h$ for some $i$'s in the range $(h+1) \leq i \leq g$. However, we can say that Theorem 2 gives a sufficiently tight lower bound on the $h$-index.

---

[2]For example, author $A$ is from *Computer Science and Engineering: Soft Computing*, author $B$ from *Computer Science and Engineering: Networks and Communication*, author $C$ is from *Electronics Engineering*, author $D$ from *Mechanical Engineering*, authors $E$ and $F$ from *Biological Sciences*.

Consider now the fifth column of Table 4 that also represents a lower bound on the $h$-index. However, this lower bound is the most loose of the three lower bounds for $h$-index. The looseness of this lower bound is due the approximation $\sum_{i=g+1}^{P} c_i \leq (P-g)h$ where we assumed that $c_i \leq h$ for $(g+1) \leq i \leq P$. Practically, $c_i << h$ for some $i$'s in the range $(g+1) \leq i \leq P$, and this accounts for the looseness of the lower bound on $h$-index.

Table 5 shows the values of $g$-index together with the values of the bounds for the $g$-index. The third column and the fourth column both represent upper bounds on the $g$-index. Specifically, the third column represents the upper bound on $g$ given by $g = O\left(\left\lceil h + \frac{e^2}{h} \right\rceil\right)$ and the fourth column represents the upper bound $g = O(h+e)$. Note that the values of the upper bound given in the fourth column are closer to the actual values of the $g$-index as compared to the values of upper bound given in the third column. This observation implies that the upper bound for $g$-index as given by Theorem 3 is more tight as compared to that given by Lemma 1. Any kind of looseness in the upper bound given in column three is due the fact that in Lemma 1, we have used an approximation $\sum_{i=h+1}^{g} c_i \leq (g-h)g$ assuming that for all $i$ in the range $(h+1) \leq i \leq g$, $c_i \leq g$. Practically, it may happen that for some $i$'s, $c_i << g$, which accounts for the looseness of the upper bound on $g$-index given by Lemma 1. In Theorem 3, the looseness of the upper bound is due to the approximation, $\sum_{i=h+1}^{g} c_i \leq (g-h)h$, assuming that for each $i$ in the range $(h+1) \leq i \leq g$, $c_i \leq g$. Practically, it may happen that for some $i$'s, $c_i << h$, which accounts for the looseness of the upper bound on $g$-index as given by Theorem 3. However, since the assumption $c_i \leq h$ is more tight than the assumption $c_i \leq g$, therefore, the upper bound given by Theorem 3 is more tight as compared to that given by Lemma 1.

Figure 1 shows $h$-index, $g$-index, $e$-index, and generalized impact factor for authors $A$ through $F$[3]. There is an observation here, although that is not our objective in this paper. The authors are arranged in the increasing order of $h$-index, we observe that all other ranking parameters are almost in the increasing order with the ascending order of $h$-index of authors.

Figure 2 shows the $h$-index and its lower bunds given by Theorem 1, Theorem 2 and Theorem 4 for authors $A$ through $F$. We observe that the lower bound given by Theorem 2 is the most tight lower bound among all three lower bounds considered in this paper. The reasons for the looseness of these lower bounds are, as explained earlier, the approximations used in deriving the expressions for these lower bounds. For example, the looseness of the lower bound on $h$ as given by Theorem 1 is due to an upper bound used $\sum_{i=h+1}^{P} c_i \leq (P-h)h$ which is based on the assumption that $c_i \leq h$ for $(h+1) \leq i \leq P$. The actual number of citations $\sum_{i=h+1}^{P} c_i$ and the corresponding upper bound, $(P-h)h$, used in Theorem 1 are shown in Figure 5. On the other hand, the lower bound on $h$-index as given by Theorem 2 uses an approximation $\sum_{i=h+1}^{g} c_i \leq (g-h)h$, which is based on the assumption that $c_i \leq h$ for $(h+1) \leq i \leq g$. The actual number of citations $\sum_{i=h+1}^{g} c_i$ and the corresponding upper bound, $(g-h)h$, used in Theorem 2 are shown in Figure 4. Further, the lower bound on $h$-index as given by Theorem 4 uses an approximation $\sum_{i=g+1}^{P} c_i \leq (P-g)h$, which is based on the assumption that $c_i \leq h$ for $(g+1) \leq i \leq P$. The actual number of citations $\sum_{i=g+1}^{P} c_i$ and the corresponding upper bound, $(P-g)h$, used in Theorem 4 are shown in Figure 6. As mentioned earlier, the lower bound given by Theorem 2 can be considered as a tight lower bound for all practical purposes.

---

[3]Authors are numbered in all these figures i.e. 1 corresponds to $A$, 2 corresponds to $B$, and so on, 6 corresponds to $F$.

Figure 3 shows the $g$-index and its upper bounds given by Lemma 1 and Theorem 3 for authors $A$ through $F$. We observe that the upper bound given by Theorem 3 is more tight as compared to the upper bound given by Lemma 1. The reasons for their looseness are, as discussed earlier, the approximations used in bounding the summations of the part of the citations that are involved in computing the lower bound. For example, the upper bound given by Theorem 3 uses an approximation $\sum_{i=h+1}^{g} c_i \leq (g-h)h$, which is based on the assumption that $c_i \leq h$ for $(h+1) \leq i \leq g$. The actual number of citations $\sum_{i=h+1}^{g} c_i$ and the corresponding upper bound, $(g-h)h$, used in Theorem 3 are shown in Figure 4. On the other hand, the upper bound on $g$-index given by Lemma 1 uses an approximation $\sum_{i=h+1}^{g} c_i \leq (g-h)g$, which is based on the assumption that $c_i \leq g$ for $(h+1) \leq i \leq g$. The actual number of citations $\sum_{i=h+1}^{g} c_i$ and the corresponding upper bound, $(g-h)g$, used in Lemma 1 are shown in Figure 4. Note that the approximation $(g-h)h$ is closer to the actual number of citations as compared to the approximation $(g-h)g$, therefore, the upper bound on $g$-index as given by Theorem 2 is more tight than the upper bound given by Lemma 1.

## 5 Conclusion

Finding the relationships among indexing parameters for determining the quality of research is a challenging task. In this paper, we describe some inequalities relating $h$-index, $g$-index, $e$-index, and generalized impact factor. We derive the inequalities from the very basic definitions of these indexing parameters and without assuming any continuous model to be followed by any of them. Also, we do not assume any prior model to be plugged in for relating these parameters. Further validation of these inequalities forms the future work.